\documentclass[12pt]{article}

\usepackage{amssymb, amsmath, amsthm, pdfpages, graphicx, natbib, bm, enumerate}
\newtheorem{theorem}{Theorem}

\newtheorem{lemma}{Lemma}

\begin{document}

\begin{center}
{\Large \bf  Minimax D-optimal designs for multivariate regression models with multi-factors}

 Lucy L. Gao$^{*}$ and Julie Zhou$^{**}$\footnote{
Corresponding author, email: jzhou@uvic.ca,
phone: 250-721-7470.}

* Department of Biostatistics \\
University of Washington,  Seattle, WA, USA 98195-7232

** Department of Mathematics and Statistics \\

University of Victoria, 
Victoria, BC, Canada V8W 2Y2 \\

\end{center}

\noindent
ABSTRACT

{\small 
In multi-response regression models, the error covariance matrix is never known in practice. Thus, there is a need for optimal designs which are robust against possible misspecification of the error covariance matrix. In this paper, we approximate the error covariance matrix with a neighbourhood of covariance matrices, in order to define minimax D-optimal designs which are robust against small departures from an assumed error covariance matrix.  It is well known that the optimization problems associated with robust designs are non-convex, which makes it challenging to construct robust designs analytically or numerically, even for one-response regression models. We show that the objective function for the minimax D-optimal design is a difference of two convex functions. This leads us to develop a flexible algorithm for computing minimax D-optimal designs, which can be applied to any multi-response model with a discrete design space. We also derive several theoretical results for minimax D-optimal designs, including scale invariance and reflection symmetry.
}



\newpage

\section{Introduction} 
\label{sec:intro} 
Consider the following multivariate regression model:
\begin{align}
&\bm y_i = Z_i^\top \bm \beta + \bm \epsilon_i, &&  i = 1, 2, \ldots, n, \label{genlm-1}\\ 
&E(\bm \epsilon_i) = \bm 0, \quad Cov(\bm \epsilon_i) = {\bf V}_\epsilon, && i = 1, 2, \ldots, n, \label{genlm-2}
\end{align} 
where $\bm y_i$ is the $i$th observed vector for the $m$ response variables $\bm y = (y_1, \ldots, y_m)^\top$, $\bm \beta = (\bm \beta_1, \ldots, \bm \beta_m)^\top$ with $\bm \beta_j \in \mathbb{R}^{q_j}$ are the $q = q_1 + \ldots + q_m$ unknown regression parameters, and $Z_i$ is given by
\begin{align} 
Z_i=\left(\begin{matrix}
{\bf f}_1^\top({\bf x}_i) & 0 & \cdots & 0 \\
0 & {\bf f}_2^\top({\bf x}_i) & \cdots & 0 \\
\vdots & \vdots & \ddots & \vdots \\
0 & 0 & \cdots & {\bf f}_m^\top({\bf x}_i) 
\end{matrix}
\right)_{m \times q}, \quad i = 1, 2, \ldots, n,
\label{Zmatrix}
\end{align} 
where ${ \bf x}_i$ is the $i$th design point for the $p$ design variables 
${\bf x} = (x_1, \ldots, x_p)^\top$ in a design space $S \subset \mathbb{R}^p$, 
and ${\bf f}_j({\bf x})$ is a $q_j$-vector of linear or non-linear functions 
of ${\bf x}$ for $j =1 , 2, \ldots, m$. The design variables ${\bf x}$ may include both quantitative variables and qualitative factors. We assume that $\bm \epsilon_i$ and $\bm \epsilon_{i'}$ are uncorrelated for $i \neq i'$. Model \eqref{genlm-1} -- \eqref{Zmatrix} is commonly used for experiments across biology, chemistry, toxicology, engineering, and other applied sciences. 

Let ${\bf W}$ be an $m \times m$ positive definite (PD) matrix. The generalized least squares estimator (GLSE) of $\bm \beta$ is given by 
\begin{align} 
\hat{\bm \beta}_{GLS} = \left ( \sum \limits_{i=1}^n Z_i^\top {\bf W}^{-1}  Z_i \right )^{-1} \left ( \sum \limits_{i=1}^n Z_i^\top {\bf W}^{-1}  \bm y_i \right ). \label{gls}
\end{align} 
Under model \eqref{genlm-1} -- \eqref{Zmatrix}, the covariance matrix of $\hat {\bm \beta}_{GLS}$ is given by
\begin{align} 
Cov(\hat{\bm \beta}_{GLS}) = \left (\sum \limits_{i=1}^n Z_i^\top {\bf W}^{-1} Z_i\right )^{-1}
\left (\sum \limits_{i=1}^n Z_i^\top {\bf W}^{-1} {\bf V}_{\epsilon} {\bf W}^{-1} Z_i\right )
\left (\sum \limits_{i=1}^n Z_i^\top {\bf W}^{-1} Z_i\right )^{-1}. \label{CovG}
\end{align} 
In the special case that ${\bf W} = \sigma^2 {\bf I}_m$  in \eqref{gls}, where $\sigma^2 > 0$ and ${\bf I}_m$ is the $m \times m$ identity matrix, the GLSE is equivalent to the ordinary least squares estimator (OLSE) of $\bm \beta$, which we denote as $\hat{\bm \beta}_{OLS}$. It follows from 
 \eqref{CovG} that 
\begin{align} 
Cov(\hat{\bm \beta}_{OLS}) = \left ( \sum \limits_{i=1}^n Z_i^\top  Z_i \right )^{-1} \left ( \sum \limits_{i=1}^n Z_i^\top  {\bf V}_\epsilon   Z_i \right ) \left ( \sum \limits_{i=1}^n Z_i^\top   Z_i \right )^{-1}. \label{CovLS}
\end{align} 

When ${\bf V}_\epsilon$ is known, we can use ${\bf W} = {\bf V}_\epsilon$ in \eqref{gls}, and the GLSE is the best linear unbiased estimator (BLUE) for $\bm \beta$. Many papers have investigated optimal designs for the GLSE with ${\bf W} = {\bf V}_\epsilon$ under model \eqref{genlm-1}--\eqref{Zmatrix}; see e.g. \citet{atashgah2007application}, \citet{atashgah2009optimal}, \citet{liu2011optimality}, \citet{liu2013note}, and \citet{wong2019optimal}. Another body of work investigated optimal designs under continuous time regression models with correlated errors \citep{dette2016optimal, dette2017new, dette2017optimal, schorning2017, dette2018optimal}, including continuous time versions of model \eqref{genlm-1}--\eqref{Zmatrix} 

Unfortunately, in practice, ${\bf V}_\epsilon$ is never known, which makes it impossible to use the GLSE with ${\bf W} = {\bf V}_\epsilon$, or the optimal designs for the GLSE with ${\bf W} = {\bf V}_\epsilon$. However, we often have a PD $m \times m$ matrix ${\bf V}_0$ which we believe is close to ${\bf V}_\epsilon$. For example, the matrix ${\bf V}_0$ may be derived from subject matter knowledge, or be derived from the results of a small pilot study. Thus, we can use ${\bf W} = {\bf V}_0$ in \eqref{gls}, or use the OLSE for $\bm \beta$. Consider the loss functions
\begin{eqnarray*}
\Phi_G(\xi,{\bf V}_0,{\bf V}_{\epsilon}) = \det\left( Cov(\hat{\bm \beta }_{GLS}) \right), \quad \Phi_L(\xi,{\bf V}_{\epsilon}) = \det\left( Cov(\hat{\bm \beta }_{OLS}) \right), 
\end{eqnarray*}
where $\xi$ represents the design measure of design points  ${\bf x}_{1}, \ldots, {\bf x}_{n}$.
We could compute D-optimal designs for the GLSE or the OLSE which minimize $\Phi_G(\xi,{\bf V}_0,{\bf V}_{\epsilon})$ or $\Phi_L(\xi,{\bf V}_{\epsilon}) $, respectively. However, the D-optimal designs would depend on the unknown ${\bf V}_\epsilon$, and computing the D-optimal designs under the assumption that ${\bf V}_\epsilon = {\bf V}_0$ could lead to a loss in efficiency when ${\bf V}_\epsilon \neq {\bf V}_0$. 

Thus, in this paper we propose a new robust minimax D-optimality criterion, which approximates ${\bf V}_\epsilon$ with a neighbourhood of matrices centred at ${\bf V}_0$. We consider both the GLSE with ${\bf W} = {\bf V}_0$ and the OLSE. The minimax approach for regression designs has been investigated in the literature to construct designs which are robust against small departures of model assumptions; see \citet{wiens2015robustness} for a review and for results for various one-response models. However, as far as the authors are aware, 
 this approach has not been studied for robust designs for multi-response models against possible misspecification of ${\bf V}_\epsilon$. 

It is extremely challenging to obtain minimax D-optimal designs analytically, even in the one-response model case, since the objective functions of the corresponding optimization problems are not convex \citep{wiens2015robustness}. Several numerical methods have been developed and used to compute optimal and robust designs, including
multiplicative, exchange, genetic, simulated annealing and particle swarm optimization algorithms.
\citet{mandal2015algorithmic} provides a review on these algorithms for finding optimal designs, and
in general they work well for convex optimization problems. \citet{atashgah2009optimal} and \citet{wong2019optimal} have also investigated efficient algorithms for solving
convex optimization problems for multivariate regression models. However, the optimization problem corresponding to the minimax D-optimal design problem is not a convex optimization problem, which makes it challenging to construct the minimax D-optimal designs numerically. 

Nevertheless, since we can show that the objective functions are differences of convex functions, we are able to use difference of convex programming (DC programming; \citealt{tao1986algorithms}, \citealt{tuy1995dc}, \citealt{lipp2016variations}, \citealt{le2018dc}) to develop a computationally efficient algorithm for computing minimax D-optimal designs on discrete design spaces.  The algorithm can be applied to find minimax D-optimal designs for any multivariate regression model with discrete design space, which  in turn makes it possible to conduct sensitivity analysis of the designs, and to explore special features of the designs. DC programming may also be very useful for solving other optimization problems in statistics. For example, \citet{nam2018nesterov} applied DC programming to a hierarchical clustering problem.

The rest of the paper is organized as follows.
In Section 2 we propose a minimax D-optimal design criterion and derive its theoretical properties.
In Section 3 we develop a general algorithm to compute minimax D-optimal designs on discrete  design spaces and obtain several results for the algorithm.
We present applications in Section 4 and make concluding remarks in Section 5.
All proofs and derivations are in the Appendix.

\section{Minimax D-optimality criterion and properties}

To deal with the unknown covariance matrix ${\bf V}_{\epsilon}$  defined in \eqref{genlm-2}, we consider approximating it with a neighbourhood (class) of matrices centred at ${\bf V}_0$, similar to a neighbourhood in \citet{wiens2008robust}:
\begin{eqnarray}
{\cal N}_{\alpha}({\bf V}_0)=\left\{{\bf V}:  ~~ {\bf V} \succeq {\bf 0} \mbox{~ and~} ||{\bf V} - {\bf V}_0|| \le \alpha \right\}, 
\label{N0}
\end{eqnarray}
where notation ``$\succeq$" denotes Loewner order for positive semi-definite matrices,
$|| \cdot||$ is any induced matrix norm, and parameter $\alpha \geq 0$ controls the  neighbourhood size.
When $\alpha=0$, ${\bf V}_0$ is the only element in ${\cal N}_{\alpha}({\bf V}_0)$. For $\alpha>0$, it can be shown \citep{wiens2008robust} that
\begin{eqnarray}
{\bf V} \preceq {\bf V}_0 + \alpha {\bf I}_m, ~\mbox{for all~} {\bf V} \in {\cal N}_{\alpha}({\bf V}_0).
\label{N0eq}
\end{eqnarray}
\citet{wiens2008robust} constructed robust designs for one-response models, while in this paper we construct robust designs for multi-response models. We focus on approximate design measures $\xi$ in the paper. 
Let the distinct support points of $\xi$ be ${\bf x}_1, \ldots, {\bf x}_k$,
and let their corresponding weights be $w_1, \ldots, w_k$ with $w_i>0$ and $\sum_{i=1}^k w_i=1$.
Define four $q \times q$ matrices,
\begin{eqnarray}
{\bf A}(\xi, {\bf V}_0) &=&  \sum_{i=1}^k w_i Z_i^\top {\bf V}_0^{-1} Z_i, \nonumber \\
{\bf B}(\xi, {\bf V}_0, {\bf V}_{\epsilon}) &=& \sum_{i=1}^k w_i Z_i^\top {\bf V}_0^{-1} {\bf V}_{\epsilon} {\bf V}_0^{-1} Z_i, \label{ABCD} \\
{\bf C}(\xi) &=& \sum_{i=1}^k w_i Z_i^\top  Z_i, \nonumber \\
{\bf D}(\xi, {\bf V}_{\epsilon}) &=& \sum_{i=1}^k w_i Z_i^\top {\bf V}_{\epsilon} Z_i. \nonumber 
\end{eqnarray}
The covariance matrices in (\ref{CovG}) and (\ref{CovLS}) are proportional to the following two matrices, respectively:
\begin{eqnarray}
{\bf M}_1(\xi, {\bf V}_0, {\bf V}_{\epsilon}) &=& 
{\bf A}^{-1}(\xi, {\bf V}_0) {\bf B}(\xi, {\bf V}_0, {\bf V}_{\epsilon}) {\bf A}^{-1}(\xi, {\bf V}_0), \label{M1M2} \\
{\bf M}_2(\xi, {\bf V}_{\epsilon}) &=& {\bf C}^{-1}(\xi) {\bf D}(\xi, {\bf V}_{\epsilon}) {\bf C}^{-1}(\xi).  \nonumber
\end{eqnarray}

We use a minimax approach to construct minimax D-optimal designs, which are robust against the misspecification of the covariance matrix ${\bf V}_{\epsilon}$. Let
\begin{eqnarray}
\phi_G(\xi, {\bf V}_0, \alpha) &=& \max_{{\bf V}_{\epsilon} \in {\cal N}_\alpha({\bf V}_0)} \log \left(
\det({\bf M}_1(\xi, {\bf V}_0, {\bf V}_{\epsilon}) ) \right), \label{Gmax}  \\
\phi_L(\xi, {\bf V}_0, \alpha) &=& \max_{{\bf V}_{\epsilon} \in {\cal N}_\alpha({\bf V}_0)} \log \left(
\det({\bf M}_2(\xi, {\bf V}_{\epsilon}) ) \right).  \nonumber
\end{eqnarray}

\noindent{\bf Definition 1:} A minimax D-optimal design based on the GLSE minimizes loss function
$\phi_G(\xi, {\bf V}_0, \alpha)$ over $\xi$ and is denoted by $\xi_G^*$.
A minimax D-optimal design based on the OLSE minimizes loss function
$\phi_L(\xi, {\bf V}_0, \alpha)$ over $\xi$ and is denoted by $\xi_L^*$. 

Various theoretical properties  of 
minimax D-optimal designs $\xi_G^*$ and $\xi_L^*$ are examined below.  First, we derive analytical formulas for $\phi_G(\xi, {\bf V}_0, \alpha)$ and $\phi_L(\xi, {\bf V}_0, \alpha)$.

\begin{theorem} 
For $ {\cal N}_\alpha({\bf V}_0) $ defined in (\ref{N0}),
\begin{eqnarray}
\phi_G(\xi, {\bf V}_0, \alpha) &=& -2 \log\left( \det({\bf A}(\xi, {\bf V}_0)) \right) +  
\log\left( \det({\bf B}(\xi, {\bf V}_0, {\bf V}_0+\alpha {\bf I}_m)) \right), \label{phiG} \\
\phi_L(\xi, {\bf V}_0, \alpha) &=& -2 \log\left( \det({\bf C}(\xi) \right) +  
\log\left( \det({\bf D}(\xi, {\bf V}_0+\alpha {\bf I}_m)) \right). \label{phiL}
\end{eqnarray} \label{Th1}
\end{theorem}
The proof of Theorem \ref{Th1} is in the Appendix.  Since we want to minimize
$\phi_G(\xi, {\bf V}_0, \alpha)$ and $\phi_L(\xi, {\bf V}_0, \alpha)$ over $\xi$ to find $\xi_G^*$ and $\xi_L^*$, respectively,
we do not need to consider any $\xi$ for which ${\bf A}(\xi, {\bf V}_0)$ or ${\bf C}(\xi)$ are singular.  Thus, in the following discussion we
only consider $\xi$ for which  ${\bf A}(\xi, {\bf V}_0)$ is nonsingular for the GLSE, or $\xi$ for which ${\bf C}(\xi)$ is nonsingular for the OLSE.  The following result shows that the matrices
${\bf B}(\xi, {\bf V}_0, {\bf V}_0+\alpha {\bf I}_m)$ and ${\bf D}(\xi, {\bf V}_0+\alpha {\bf I}_m)$ are
also nonsingular if ${\bf A}(\xi, {\bf V}_0)$ and ${\bf C}(\xi)$ are nonsingular, respectively.

\begin{lemma} 
If 	${\bf A}(\xi, {\bf V}_0)$ is nonsingular, then ${\bf B}(\xi, {\bf V}_0, {\bf V}_0+\alpha {\bf I}_m)$ is nonsingular for all $\alpha \ge 0$.
If  ${\bf C}(\xi)$ is nonsingular, then ${\bf D}(\xi, {\bf V}_0+\alpha {\bf I}_m)$ is nonsingular for all $\alpha > 0$.
\label{Lemma1}
\end{lemma}	

The proof of Lemma \ref{Lemma1} is in the Appendix. Next, we consider the convexity of 
$\phi_G(\xi, {\bf V}_0, \alpha)$ and $\phi_L(\xi, {\bf V}_0, \alpha)$ as a function of $\xi$. 
Suppose there are two design measures $\xi_1$ and $\xi_2$ having the same support points ${\bf x}_1, \ldots, {\bf x}_k$, but with different weights.
Let $w_1^j, \ldots, w_k^j$ be the weights for $\xi_j$, $j=1$ and 2.
We define a convex combination of $\xi_1$ and $\xi_2$ to be 
$\xi_{\delta}=(1-\delta)\xi_1+\delta ~\xi_2$, where $\xi_\delta$ has the same support points as $\xi_1$ and $\xi_2$, and the weights are given by $(1-\delta)w_1^1+\delta ~w_1^2, \ldots, (1-\delta)w_k^1+\delta~ w_k^2$, where $\delta \in [0, 1]$.

\begin{theorem} 
	For fixed ${\bf V}_0$  and $\alpha > 0$, 
	$\phi_G(\xi_{\delta}, {\bf V}_0, \alpha)$ is a difference of two convex functions of $\delta$, and so is
	$\phi_L(\xi_{\delta}, {\bf V}_0, \alpha)$.
	\label{Th2}
\end{theorem}	

The proof of Theorem \ref{Th2} is in the Appendix. 
It is well known that robust design loss functions are not convex functions 
in terms of $\xi$,  which makes it challenging to compute robust designs. 
However, the result in Theorem \ref{Th2} provides useful information about $ \phi_G(\xi_{\delta}, {\bf V}_0, \alpha)$
and $\phi_L(\xi_{\delta}, {\bf V}_0, \alpha) $, which allows us to develop an efficient and 
effective algorithm in Section 3.  

Now we investigate scale invariance and other properties of minimax D-optimal designs.
Consider a design space $S$ for model \eqref{genlm-1} -- \eqref{Zmatrix} and its scale transformation $T$, say
$T{\bf x}=(t_1x_1, \ldots, t_px_p)^\top$, where $t_1, \ldots, t_p$ are positive numbers.
Let $S_T$ denote the transformed design space, i.e., $S_T=\{ T{\bf x}: ~ {\bf x} \in S\}$.

\noindent{\bf Definition 2:} 
Suppose $\xi^*$ is a minimax D-optimal design on $S$ based on the GLSE or OLSE, with 
support points ${\bf x}_1^*, \ldots, {\bf x}_k^*$ and corresponding weights
$w_1^*, \ldots, w_k^*$.  We say $\xi^*$ is scale invariant if the design with
support points $T{\bf x}_1^*, \ldots, T{\bf x}_k^*$ and corresponding weights
$w_1^*, \ldots, w_k^*$ is a minimax D-optimal design on $S_T$. 

Minimax D-optimal designs are scale invariant for some multivariate regression models.
Theorem \ref{Th3} below provides a sufficient condition to check for the scale invariance
of $\xi_G^*$ and $\xi_L^*$.

\begin{theorem} 
If the vectors ${\bf f}_1({\bf x}), \ldots, {\bf f}_m({\bf x})$ in model \eqref{genlm-1} -- \eqref{Zmatrix} satisfy the following condition,
for $j=1, \ldots, m$,
$${\bf f}_j(T{\bf x}) = Q_j {\bf f}_j({\bf x}), ~\mbox{for all~} {\bf x} \in S,$$
where each ${Q}_j$ is a diagonal matrix and the diagonal elements do not depend on ${\bf x}$,
then both $\xi_G^*$ and $\xi_L^*$ are scale invariant.
\label{Th3}
\end{theorem}

The proof of Theorem \ref{Th3} is in the Appendix. 
The scale invariance property allows us to find minimax D-optimal designs on the scaled design space, which
can reduce the computation time if we need to construct minimax D-optimal designs for several design spaces which differ only in size. 

Minimax D-optimal designs $\xi_G^*$ and $\xi_L^*$ usually depend on ${\bf V}_0$, but they may depend on the covariances in ${\bf V}_0$  through their absolute values.
For instance, when $m=2$, let
\begin{eqnarray}
 {\bf V}_0 =\left( \begin{array}{cc}
\sigma_1^2  & \sigma_{12} \\
\sigma_{12} & \sigma_2^2
\end{array}  \right),  ~~~~~
{\bf V}_1 =\left( \begin{array}{cc}
\sigma_1^2  & -\sigma_{12} \\
-\sigma_{12} & \sigma_2^2
\end{array}  \right). 
\label{V0V1} 
\end{eqnarray}
Then, using ${\bf V}_1$  in (\ref{phiG}) and (\ref{phiL}) leads to the same minimax D-optimal designs $\xi_G^*$ and $\xi_L^*$ 
as those from using ${\bf V}_0$, which indicates that $\xi_G^*$ and $\xi_L^*$ only depend on the absolute value of $\sigma_{12}$.
This result can be proved from a general result that we derive in the next theorem.

\begin{theorem} 
	Suppose ${\bf V}_1$ is an $m \times m$ covariance matrix.   If there exists a diagonal matrix ${\bf Q}$ 
	with the diagonal elements taking two possible values
	$+1$ and $-1$	such that 
	${\bf V}_1 = {\bf Q} {\bf V}_0 {\bf Q}$, then
	using ${\bf V}_1$  in (\ref{phiG}) and (\ref{phiL}) leads to the same minimax D-optimal designs $\xi_G^*$ and $\xi_L^*$ 
	as those when  ${\bf V}_0$ is used.
	\label{Th4}	
\end{theorem}

The proof of Theorem \ref{Th4} is in the Appendix.  This result does not depend on the vectors
${\bf f}_1({\bf x}), \ldots, {\bf f}_m({\bf x})$, so it is true for any multivariate model.
When $m=2$, it is easy to show  that the two diagonal elements of  ${\bf Q}$ are $+1$ and $-1$ and
${\bf V}_1 = {\bf Q} {\bf V}_0 {\bf Q}$ holds for the matrices in (\ref{V0V1}).
When $m=3$, for instance we can show that the following ${\bf V}_0$ and ${\bf V}_1$ satisfy the condition in 
Theorem \ref{Th4} and hence yield the the same minimax D-optimal designs:
\begin{eqnarray}
{\bf V}_0 =\left( \begin{array}{ccc}
\sigma_1^2  & \sigma_{12}  & 0 \\
\sigma_{12} & \sigma_2^2 & \sigma_{23} \\
0 & \sigma_{23} & \sigma_3^2  
\end{array}  \right),  ~~~~~
{\bf V}_1 =\left( \begin{array}{ccc}
\sigma_1^2  & -\sigma_{12}  & 0 \\
-\sigma_{12} & \sigma_2^2 & -\sigma_{23} \\
0 & -\sigma_{23} & \sigma_3^2  
\end{array}  \right). 
\nonumber
\end{eqnarray}
There are other matrices that yield the the same minimax D-optimal designs for $m=3$; see Example 1 in Section 4 for a demonstration.  The above result can also be generalized and applied for $m>3$ easily.

When the vectors  in model \eqref{genlm-1} -- \eqref{Zmatrix} are equal, i.e., 
${\bf f}_1({\bf x})= \ldots= {\bf f}_m({\bf x})$,
$\xi_G^*$ and $\xi_L^*$  do not depend on ${\bf V}_0$ and $\alpha$.
In fact $\xi_G^*$ and $\xi_L^*$ are the same as those D-optimal designs for 
model \eqref{genlm-1} -- \eqref{Zmatrix} with $m=1$.
This result can be proved using Lemma 2 in \citet{wong2019optimal}.
In addition, if $\xi_G^*$ and $\xi_L^*$  do not depend on ${\bf V}_0$, then
$\xi_G^*$ and $\xi_L^*$ are the same.  This is due to the fact that, from (\ref{ABCD}), ${\bf V}_0 = {\bf I}_m$ gives 
$$ {\bf A}(\xi, {\bf I}_m) =  {\bf C}(\xi), \quad  
{\bf B}(\xi, {\bf I}_m, {\bf V}_\epsilon)= {\bf D}(\xi, {\bf V}_\epsilon), \quad \mbox{for any} ~\xi.$$

After discussing a numerical algorithm for finding minimax D-optimal designs in Section 3, we can derive 
more theoretical results for $\xi_G^*$ and $\xi_L^*$.

\section{Numerical method}

We develop a general algorithm to compute minimax D-optimal designs on discrete design spaces.
Let $S_N=\{ {\bf u}_1, \ldots, {\bf u}_N \}  \subset \mathbb{R}^p$ denote a discrete design space with $N$ points,
where points ${\bf u}_1, \ldots, {\bf u}_N$ are user selected. For any compact design space $S$, we construct $S_N$ by including a large number of grid points to cover $S$. 

For any $\xi$ on  $S_N$, let weight vector ${\bf w}=(w_1, \ldots, w_N)^\top$ contain the weights for
all the points in $S_N$ with $w_i$ being the weight at point ${\bf u}_i$.
These weights satisfy  $w_i \ge 0$ and $\sum_{i=1}^N w_i=1$.
If a point receives a positive weight, then the point becomes a support point of $\xi$.
To state the minimax D-optimal design problems on $S_N$, we introduce matrices
\begin{eqnarray}
\tilde{\bf A}({\bf w}, {\bf V}_0) &=&  \sum_{i=1}^N w_i Z_i^\top {\bf V}_0^{-1} Z_i, \nonumber \\
\tilde{\bf B}({\bf w}, {\bf V}_0, {\bf V}_0+\alpha {\bf I}_m) &=& 
\sum_{i=1}^N w_i Z_i^\top {\bf V}_0^{-1} ({\bf V}_0+\alpha {\bf I}_m) {\bf V}_0^{-1} Z_i, \label{ABCD2} \\
\tilde{\bf C}({\bf w}) &=& \sum_{i=1}^N w_i Z_i^\top  Z_i, \nonumber \\
\tilde{\bf D}({\bf w}, {\bf V}_0+\alpha {\bf I}_m) &=& \sum_{i=1}^N w_i Z_i^\top ({\bf V}_0+\alpha {\bf I}_m) Z_i, \nonumber 
\end{eqnarray}
where matrices $Z_i$, defined in (\ref{Zmatrix}), are now evaluated at
${\bf f}_1^\top({\bf u}_i), \ldots, {\bf f}_m^\top({\bf u}_i)$ for $i=1, \ldots,N$.
Define loss functions
\begin{eqnarray}
&& \hspace{-1cm} \tilde{\phi}_G({\bf w}, {\bf V}_0, \alpha) = -2 \log\left( \det(\tilde{\bf A}({\bf w}, {\bf V}_0)) \right) +  
\log\left( \det(\tilde{\bf B}({\bf w}, {\bf V}_0, {\bf V}_0+\alpha {\bf I}_m)) \right), \label{phiG2} \\
&& \hspace{-1cm} \tilde{\phi}_L({\bf w}, {\bf V}_0, \alpha) = -2 \log\left( \det(\tilde{\bf C}({\bf w}) \right) +  
\log\left( \det(\tilde{\bf D}({\bf w}, {\bf V}_0+\alpha {\bf I}_m)) \right). \label{phiL2}
\end{eqnarray}

From Theorem \ref{Th1}, the minimax D-optimal designs on $S_N$  based on the GLSE and OLSE minimize
$\tilde{\phi}_G({\bf w}, {\bf V}_0, \alpha)$
and $\tilde{\phi}_L({\bf w}, {\bf V}_0, \alpha)$ over ${\bf w}$,
respectively.
By (\ref{ABCD2}) matrices 
$\tilde{\bf A}({\bf w}, {\bf V}_0)$,
$\tilde{\bf B}({\bf w}, {\bf V}_0, {\bf V}_0+\alpha {\bf I}_m)$,
$\tilde{\bf C}({\bf w})$, and
$\tilde{\bf D}({\bf w}, {\bf V}_0+\alpha {\bf I}_m)$
are all linear in ${\bf w} $.
Similar to the proof of Theorem \ref{Th2}, we can show that 
for fixed ${\bf V}_0$ and $\alpha$, loss function
$\tilde{\phi}_G({\bf w}, {\bf V}_0, \alpha)$
or $\tilde{\phi}_L({\bf w}, {\bf V}_0, \alpha)$  is a
difference of convex functions of ${\bf w} $.

A general  minimax D-optimal design problem on $S_N$ can then be written as
\begin{eqnarray}
&& \min_{\bf w}  g({\bf w}) - h({\bf w}) \label{DC1} \\
&& \mbox{subject to:} ~w_i \ge 0, ~\mbox{for}~i=1,\ldots, N,~\sum_{i=1}^Nw_i=1, \nonumber
\end{eqnarray}
where both $g({\bf w})$ and $ h({\bf w})$ are convex functions of ${\bf w} $.
For the loss functions in (\ref{phiG2}) and (\ref{phiL2}) it is easy to write out the 
corresponding functions
$g({\bf w})$ and $ h({\bf w})$.
Let $\nabla h({\bf w})$ be the gradient vector of $ h({\bf w})$; the closed form expression for $\nabla h({\bf w})$ can be found in the Appendix.  
Let $v({\bf w}, {\bf w}^0)=h({\bf w}^0)+ ({\bf w}- {\bf w}^0)^\top \nabla h({\bf w}^0)$ be
the first order approximation of $ h({\bf w})$ at point ${\bf w}^0$.
The key to solving problem (\ref{DC1}) is to work on a closely related problem as follows:
for a given ${\bf w}^0$,
\begin{eqnarray}
&& \min_{\bf w}  g({\bf w}) - v({\bf w}, {\bf w}^0) \label{DC1A} \\
&& \mbox{subject to:} ~w_i \ge 0, ~\mbox{for}~i=1,\ldots, N,~\sum_{i=1}^Nw_i=1. \nonumber
\end{eqnarray} 
The difference between (\ref{DC1})  and (\ref{DC1A}) is in the objective function. In particular,  the 
objective function in (\ref{DC1A}) is convex. 

We propose an iterative algorithm to solve problem (\ref{DC1}). The details are provided in Algorithm 1.

\noindent
\noindent{\underline{\bf Algorithm 1 \hspace{12cm}}}
\begin{description}
	\item [Step 1: Initialization]  
	For a given model and design space $S_N$, compute matrices $Z_i$, $i=1, \ldots, N$.  Set up the values of $\alpha$ and
	${\bf V}_0$.  Choose an initial weight vector ${\bf w}^{(0)}$ such that 
	$\tilde{\bf A}({\bf w}^{(0)}, {\bf V}_0)$ or $\tilde{\bf C}({\bf w}^{(0)})$ is nonsingular, depending on the
	design problem to be solved.
	
	\item [Step 2: Iteration]  For $l=1, 2, \ldots$, repeat the following until $\|{\bf w}^{(l)} - {\bf w}^{(l-1)}\| < \eta_1$
	for a small positive $\eta_1$: 
	
	Solve problem (\ref{DC1A}) using ${\bf w}^0 = {\bf w}^{(l-1)}$ in $v({\bf w}, {\bf w}^0)$
	and denote the solution as ${\bf w}^{(l)}$.
		
\end{description}
\vspace{-0.5cm}
\noindent{\underline{\hspace{15cm}}}

Let ${\bf w}^{(l)}$ be the weight vector after iteration $l$.  We define convergence in Algorithm 1 as
$|| {\bf w}^{(l)} - {\bf w}^{(l-1)}|| < \eta_1$ for a small positive $\eta_1$.
The limit ${\bf w}^{*}$ of ${\bf w}^{(l)}$ as $l \to \infty$ is a solution to problem (\ref{DC1}),
which gives a minimax D-optimal design.

\noindent{Remarks:}
\begin{description}
\item (i)
Problem (\ref{DC1A}) is a convex optimization problem and there are  efficient algorithms to solve it.
CVX program in MATLAB has been used successfully to solve convex optimization problems for
finding various optimal regression designs;  for example, see
\citet{wong2019optimal} for many applications and properties of CVX program.
Thus, in Step 2  we can apply CVX program to find ${\bf w}^{(l)}$ easily and we also know that 
CVX program can solve the problem with large $N$.

\item (ii)  We can get an initial weight vector ${\bf w}^{(0)}$	from the solution of problem (\ref{DC1})
by replacing the objective function  with  $g({\bf w})$.  Since $g({\bf w})$ is a convex function of ${\bf w}$,
CVX can be applied to find the solution.  This  initial weight vector 
guarantees that $\tilde{\bf A}({\bf w}^{(0)}, {\bf V}_0)$ or 
$\tilde{\bf C }({\bf w}^{(0)})$ is nonsingular, and it
works well for all the examples in this paper.

\item (iii)
If the sequence ${\bf w}^{(l)}$ converges to a weight vector, say ${\bf w}^{*}$, as $l \to \infty$,
then $$\lim_{l \to \infty} g({\bf w}^{(l)}) - v({\bf w}^{(l)}, {\bf w}^{(l-1)})
	=\lim_{l \to \infty} g({\bf w}^{(l)}) - h({\bf w}^{(l)}) = g({\bf w}^{*}) - h({\bf w}^{*}).$$

\item (iv) The gradient vectors of $g({\bf w}) - v({\bf w}, {\bf w}^{(l-1)})$
and  $g({\bf w} ) - h({\bf w} )$, evaluated at ${\bf w}^{(l)}$, converge to the same limit
$\nabla g({\bf w}^{*})- \nabla h({\bf w}^{*})$ as $l \to \infty$.
\end{description}

By Remarks (iii) and (iv), the limiting weight vector ${\bf w}^{*}$ should satisfy the first order condition as
a local minimizer of problem (\ref{DC1}). Alternatively,
we can derive the optimality condition of the local minimizer from design theory as follows.

\begin{theorem} 
For fixed $\alpha$ and ${\bf V}_0$ the local minimizer ${\bf w}^{*}$ of  problem (\ref{DC1})
with objective (loss) functions in (\ref{phiG2}) and (\ref{phiL2}) must
satisfy,  for $i=1, \ldots, N$,
$$  \mbox{tr} \left( 2 {\bf G}^{-1}({\bf w}^{*}) G_i - {\bf H}^{-1}({\bf w}^{*}) H_i \right) - q \le 0,$$
where 
\begin{eqnarray*}
&&G_i=\left\{ \begin{array}{ll}
	Z_i^\top {\bf V}_0^{-1} Z_i, & \mbox{for loss function in ~} (\ref{phiG2}), \\
	Z_i^\top  Z_i, & \mbox{for loss function in ~} (\ref{phiL2}), 
\end{array}
\right.  \\
&& H_i=\left\{ \begin{array}{ll}
	Z_i^\top {\bf V}_0^{-1} ({\bf V}_0 + \alpha {\bf I}_m) {\bf V}_0^{-1} Z_i, & \mbox{for loss function in~} (\ref{phiG2}), \\
	Z_i^\top ({\bf V}_0 + \alpha {\bf I}_m) Z_i, & \mbox{for loss function in ~} (\ref{phiL2}), 
\end{array}
\right.  \\
&& {\bf G}({\bf w})	= \sum_{i=1}^N w_i G_i ~~~\mbox{and}~~~ {\bf H}({\bf w})	= \sum_{i=1}^N w_i H_i.
\end{eqnarray*}
	\label{Th5}
\end{theorem}

The proof of Theorem \ref{Th5} is in the Appendix.  In practice, we relax the condition in
Theorem \ref{Th5} to
\begin{eqnarray}
\mbox{tr} \left( 2 {\bf G}^{-1}({\bf w}^{*}) G_i - {\bf H}^{-1}({\bf w}^{*}) H_i \right) - q \le \eta_2, ~~	
\mbox{for~} i=1, \ldots, N,
\label{Cont2}
\end{eqnarray}
where $\eta_2$ is a small positive number.  We use this condition to verify that numerical results from Algorithm 1 are minimax D-optimal designs.

From Algorithm 1 we can investigate reflection symmetry of $\xi_G^*$ and $\xi_L^*$.
When $S_N$ has reflection symmetry, $\xi_G^*$ and $\xi_L^*$ also have this property for some models.
We obtain a sufficient condition to check for this property below.
Let $T_r$ be a reflection transformation with respect to variable $x_r$, i.e.,
$T_r {\bf x} =(x_1, \ldots, x_{r-1}, -x_r, x_{r+1}, \ldots, x_p)^\top$. Define
$S_{T_r} =\{ T_r {\bf x}:  {\bf x} \in S_N\}$.  If $S_N = S_{T_r}$, then 
$S_N$ has reflection symmetry with respect to variable $x_r$.

\begin{theorem}
Suppose $S_N$ has reflection symmetry with respect to variable $x_r$.  
If the vectors ${\bf f}_1({\bf x}), \ldots, {\bf f}_m({\bf x})$ in model \eqref{genlm-1} -- \eqref{Zmatrix} satisfy the following condition,
for $j=1, \ldots, m$,
$${\bf f}_j(T_r{\bf x}) = Q_j {\bf f}_j({\bf x}), ~\mbox{for all~} {\bf x} \in S_N,$$
where each ${Q}_j$ is a diagonal matrix and the diagonal elements are constants being either $+1$ or $-1$,
then there exist  $\xi_G^*$ and $\xi_L^*$ that have reflection symmetry with respect to variable $x_r$.
\label{Th6}
\end{theorem}
The proof of Theorem \ref{Th6} is in the Appendix.  
The reflection symmetry property of $\xi_G^*$ and $\xi_L^*$
 can be applied sequentially for several design variables if $S_N$ has the property.
When $N$ is large, the result in Theorem \ref{Th6}  is very useful to reduce the computation 
time for finding $\xi_G^*$ and $\xi_L^*$, by reducing the number of unknown weights $w_i$ in Algorithm 1.

\section{Applications}

We present three examples to construct $\xi_G^*$ and $\xi_L^*$ using Algorithm 1.
In Example 1 there are both quantitative and qualitative factors, and  the design space contains $N=4400$ points.  We demonstrate that the reflection symmetry property can greatly reduce the computation time in Algorithm 1, and that Algorithm 1 can accurately and quickly find optimal designs.  Various properties of  
$\xi_G^*$ and $\xi_L^*$ are discussed as well.
Example 2 considers multivariate regression with quadratic and cubic spline functions, and 
Algorithm 1 is flexible to find $\xi_G^*$ and $\xi_L^*$ easily.  
This allows us to find interesting features of $\xi_G^*$ and $\xi_L^*$.
In Example 3  another property of $\xi_G^*$ and $\xi_L^*$ is explored.  In particular, we find a case
where $\xi_G^*$ and $\xi_L^*$ do not depend on ${\bf V}_0$ and $\alpha$. 

We have used MATLAB software to implement Algorithm 1, since the CVX program in MATLAB is very fast.
The MATLAB code for all the examples in this paper is available from the authors upon request.
All the compuation is done on a PC equipped with 
Intel Core i7-8700 Six Core 4.6 GHz CPU 16 GB 2666 MHz DDR4.
In Algorithm 1 we set $\eta_1=10^{-5}$ in the stopping criterion, and we also use
the condition in (\ref{Cont2}) to verify for minimax D-optimal designs.

\noindent{\bf Example 1.} Consider model \eqref{genlm-1} -- \eqref{Zmatrix} with $m=3$ and $p=5$ design
variables, and 
\begin{eqnarray*}
	{\bf f}_1({\bf x})	&=&  (1, x_1, x_2, x_3,x_4,x_5,x_1x_4, x_1x_5, x_2x_4, x_2x_5, x_3x_4, x_3x_5)^\top, \\
	{\bf f}_2({\bf x})	&=&  (1, x_1, x_2, x_3,x_4,x_5,x_1x_3^3, x_4x_3^2)^\top, \\
	{\bf f}_3({\bf x})	&=&  (1, x_1, x_2, x_3,x_4,x_5, x_3^3)^\top,
\end{eqnarray*}
where $x_1 \in [-1, 1],  x_2 \in [-1, 1]$ and $x_3 \in [-2, 2]$  are quantitative variables, 
and $x_4 = 0, 1$ and $x_5=0, 1$ are qualitative variables.
We compute minimax D-optimal designs on $S_N$ with $N=4N_1N_2N_3$, where $N_j$ equally spaced grid points are used for each
$x_j$, $j=1,2,3$. We use $N_1=10, N_2=10$ and $N_3=11$ to illustrate the computation and properties of $\xi_G^*$ and $\xi_L^*$.
By Theorem \ref{Th6},  there exist $\xi_G^*$ and $\xi_L^*$ that have reflection symmetry with respect to $x_1$, $x_2$ and $x_3$ for
the model on $S_N$.  Since $N_3$ is odd, we just use the symmetry with respect to  $x_1$ and $x_2$ to reduce the unknown weights to
$N/4=1100$ in Algorithm 1. Let  
$${\bf V}_0 = \left(  \begin{array}{ccc}
3 & -1 & 0 \\
-1 & 9 & 6 \\
0 & 6 & 16 
\end{array}  \right). $$
Representative $\xi_G^*$ and $\xi_L^*$ are given in Table \ref{table1} for $\alpha=0, 3, 8$ and 10. The results indicate that Algorithm 1 is effective and efficient; it takes between 75 to 547 seconds to find $\xi_G^*$ and $\xi_L^*$ for $N=4400$ and $q=27$.  
When the initial weight ${\bf w}^{(0)}$ is closer to the solution, it takes less computation time.

The support points of  $\xi_G^*$ and $\xi_L^*$ for all the cases are the same for this model, but the weights are slightly different.
As expected, the loss function $\tilde{\phi}_G$ is smaller than $\tilde{\phi}_L$ for $\alpha \le 7$, and 
$\tilde{\phi}_G$ is larger than $\tilde{\phi}_L$ for $\alpha \ge 8$.
This  implies that the GLSE is more efficient than the OLSE if ${\bf V}_{\epsilon}$ is in a smaller neighbourhood of ${\bf V}_0$.
We have also computed $\xi_G^*$ and $\xi_L^*$  when ${\bf V}_0$ is replaced by one of the following matrices:
$$\left(  \begin{array}{ccc}
3 & 1 & 0 \\
1 & 9 & 6 \\
0 & 6 & 16 
\end{array}  \right),
\quad
\left(  \begin{array}{ccc}
3 & -1 & 0 \\
-1 & 9 & -6 \\
0 & -6 & 16 
\end{array}  \right), 
\quad
\left(  \begin{array}{ccc}
3 & 1 & 0 \\
1 & 9 & -6 \\
0 & -6 & 16 
\end{array}  \right).
$$
We obtain the same $\xi_G^*$ and $\xi_L^*$ as in Table \ref{table1}, which confirms the result in Theorem \ref{Th4}.
The scale invariance result in Theorem \ref{Th3} is also true for  $\xi_G^*$ and $\xi_L^*$ and we verified it with our numerical results.

\begin{table}
\begin{small}
	\caption{Minimax D-optimal designs in Example 1. For each design only $1/4$ of the support points and weights are listed, and the other
		$3/4$ of the support points and weights can be obtained  by the reflection symmetry with respect to  $x_1$ and $x_2$.}
\begin{center}	
\begin{tabular}{lcccc} \hline
	support points &  \multicolumn{4}{c}{weights for $\xi_G^*$ (and $\xi_L^*$ in parentheses) } \\
	$(x_1,x_2,x_3,x_4,x_5)$ & $\alpha=0$ & $\alpha=3$  & $\alpha=8$  & $\alpha=10$  \\ \hline
	$(1,1,2,0,0) $& .0239 (.0252) & .0242 (.0248) & .0246 (.0246) & .0247 (.0246) \\
	$(1,1,2,0,1) $& .0239 (.0252) & .0242 (.0248) & .0246 (.0246) & .0247 (.0246) \\
	$(1,1,2,1,0) $& .0224 (.0221) & .0223 (.0222) & .0222 (.0222) & .0222 (.0222) \\
	$(1,1,2,1,1) $& .0224 (.0221) & .0223 (.0222) & .0222 (.0222) & .0222 (.0222) \\
	$(1,1,0,0,0) $& .0102 (.0054) & .0090 (.0067) & .0076 (.0075) & .0071 (.0076) \\
	$(1,1,0,0,1) $& .0102 (.0054) & .0090 (.0067) & .0076 (.0075) & .0071 (.0076) \\
	$(1,1,0,1,0) $& .0222 (.0250) & .0230 (.0243) & .0238 (.0239) & .0241 (.0238) \\
	$(1,1,0,1,1) $& .0222 (.0250) & .0230 (.0243) & .0238 (.0239) & .0241 (.0238) \\
	$(1,1,-2,0,0) $& .0239 (.0252) & .0242 (.0248) & .0246 (.0246) & .0247 (.0246) \\
	$(1,1,-2,0,1) $& .0239 (.0252) & .0242 (.0248) & .0246 (.0246) & .0247 (.0246) \\
	$(1,1,-2,1,0) $& .0224 (.0221) & .0223 (.0222) & .0222 (.0222) & .0222 (.0222) \\
	$(1,1,-2,1,1) $& .0224 (.0221) & .0223 (.0222) & .0222 (.0222) & .0222 (.0222) \\ \hline
	loss function $\tilde{\phi}_G$ & $55.4173$  & $68.7782$   & $81.4346$   & $85.0921$ \\
~~~~~~~~~~~~~~~~~$\tilde{\phi}_L$ & $56.3063$  & $69.1105$   & $81.4025$   & $84.9781$ \\ \hline
computation $\xi_G^*$  & 74.1719 & 355.8281 & 502.2969 & 546.6719 \\
time (s):  ~~~~~$\xi_L^*$& 266.4531 & 210.2344 & 162.5000 & 144.1875 \\	
\hline
\end{tabular}		
\end{center}
\label{table1}
\end{small}	
\end{table}

\noindent{\bf Example 2.} Consider model \eqref{genlm-1} -- \eqref{Zmatrix} with $m=3$ and 2 design
variables, and 
\begin{eqnarray*}
{\bf f}_1({\bf x})	&=&  (1, x_1, x_2, x_1x_2, x_1^2, x_2^2)^\top, \\
{\bf f}_2({\bf x})	&=&  (1, x_1, x_1^2, x_1^3, (x_1-0.5)_+^3, (x_1+0.5)_+^3)^\top, \\
{\bf f}_3({\bf x})	&=&  (1, x_2, x_2^2)^\top,
\end{eqnarray*}
where function $(s)_+ = \max(0,s)$. The three expected responses include quadratic and cubic spline functions, and there are $q=15$ regression parameters. The design space $S_N$ contains $N=21^2$ grid points in
$[-1, +1]^2$, with both $x_1$ and $x_2$ taking 21 equally spaced values $-1, -0.9, -0.8, \ldots, +0.8, +0.9, +1$.
Let  $${\bf V}_0=\left( \begin{array}{ccc}
4 & 3 & 4 \\
3 & 9 & 6 \\
4 & 6 & 16 
\end{array} \right). $$
Using Algorithm 1 we compute $\xi_G^*$ and $\xi_L^*$ for various $\alpha$ values.
Representative results and computation times are given in Table \ref{table2}.

\begin{figure}[ht]
	\centering
	\includegraphics[width=6.0in]{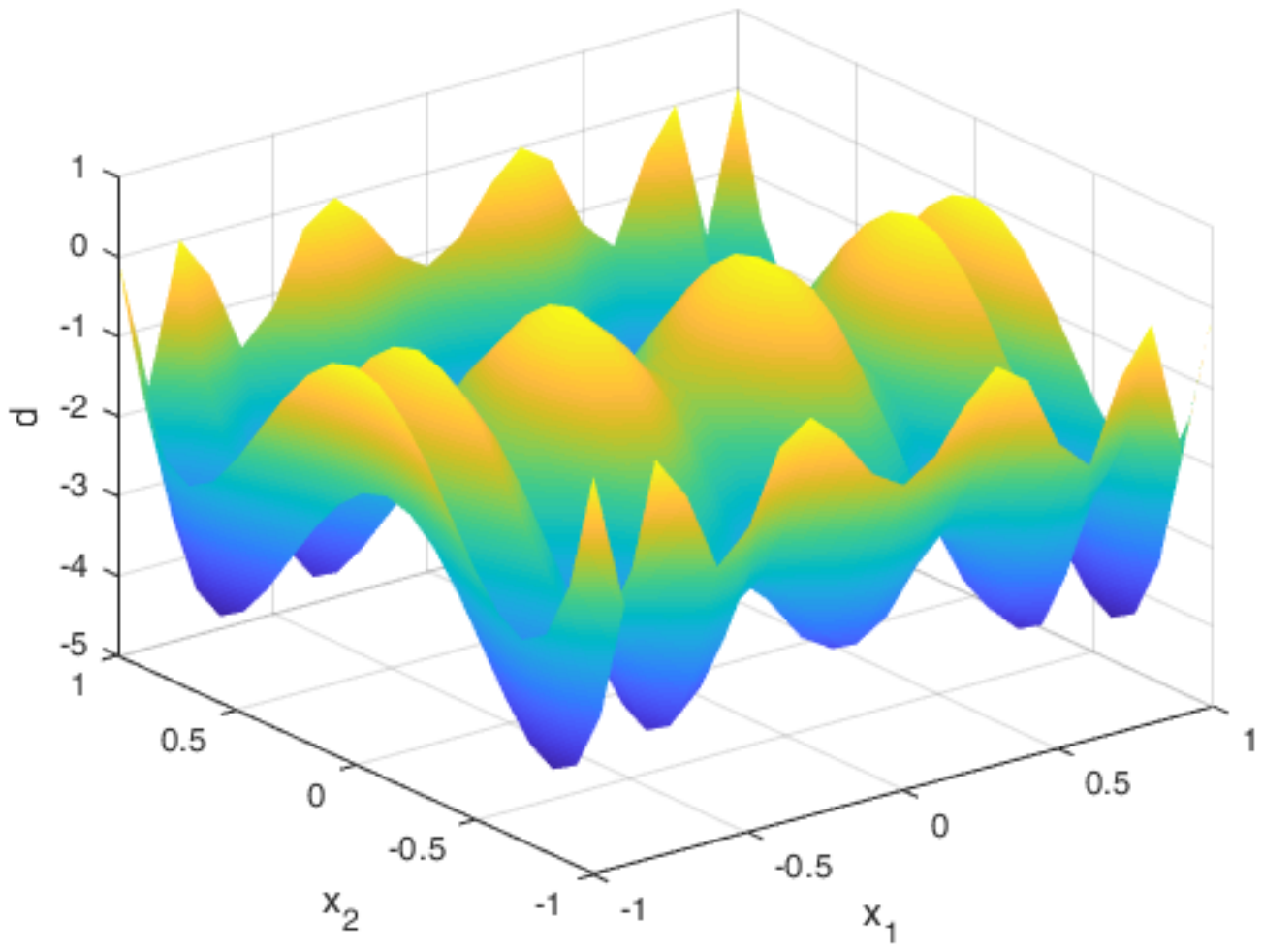}
	\caption{Plot of $d(x_1,x_2)$ versus $(x_1,x_2) $ ($\in S_{N}$)   for $\xi_G^*$ with $\alpha=5.0$}
	\label{Fig1}
\end{figure}

\begin{figure}[ht]
	\centering
	\includegraphics[width=6.0in]{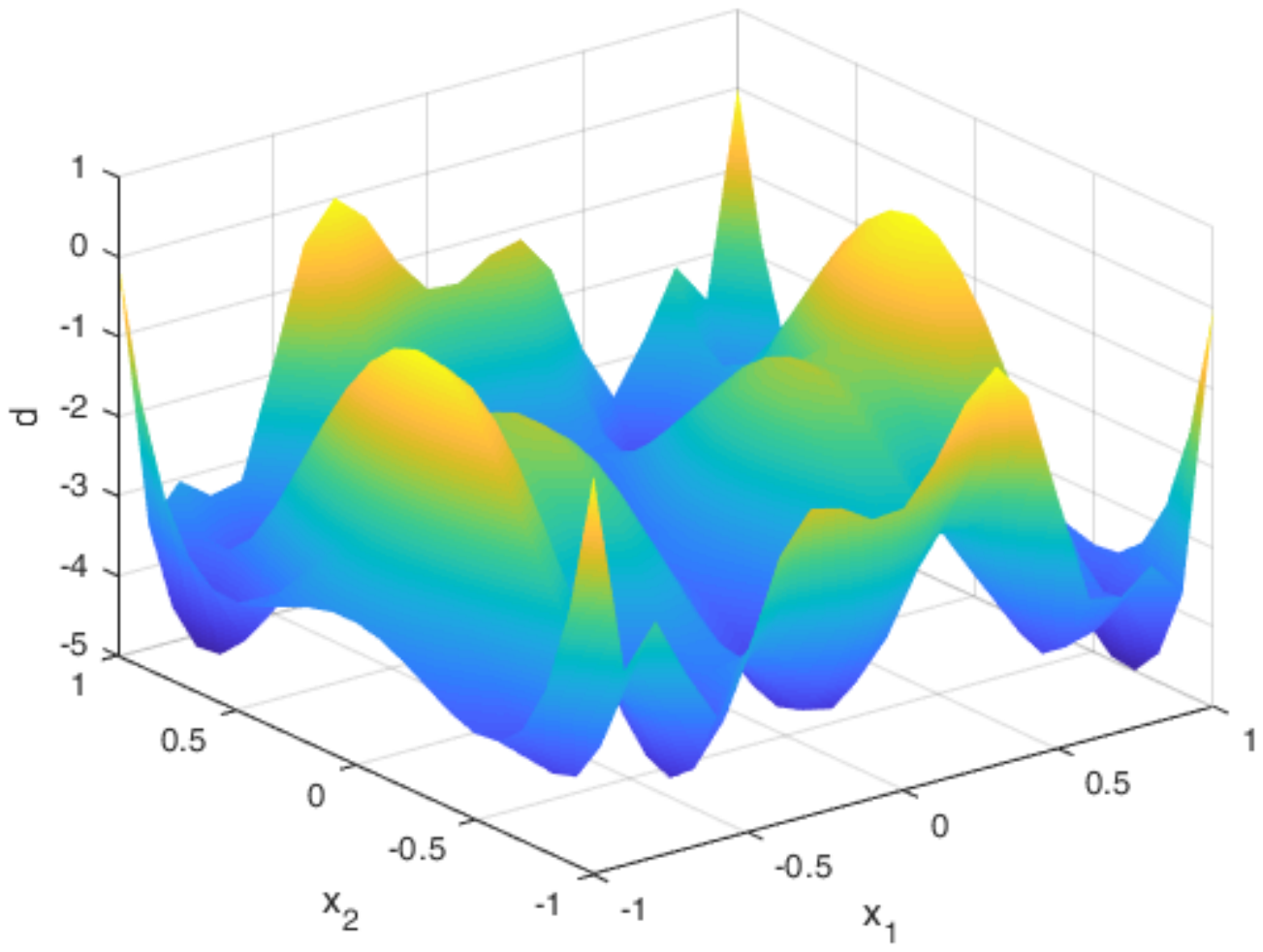}
	\caption{Plot of $d(x_1,x_2)$ versus $(x_1,x_2) $ ($\in S_{N}$)   for $\xi_L^*$ with $\alpha=0.0$}
	\label{Fig2}
\end{figure}

Since we can use Algorithm 1 to find 
$\xi_G^*$ and $\xi_L^*$ for various situations, we can easily study the features in $\xi_G^*$ and $\xi_L^*$:
\begin{enumerate}[(1)]
\item 
$\xi_G^*$ is less sensitive to small changes in $\alpha$ than $\xi_L^*$.  The GLSE is more efficient than the
OLSE at the minimax designs for small $\alpha$, since $ \tilde{\phi}_G < \tilde{\phi}_L$.
Notice that Table \ref{table2} only shows $1/2$ of the support points in $\xi_G^*$.
For $\alpha=0, 3$ and 5, there are fewer support points in $\xi_L^*$ than in $\xi_G^*$.

\item We have used all the points in $S_N$ to find $\xi_G^*$ and $\xi_L^*$.
$\xi_G^*$  shows the reflection symmetry with respect to $x_1$ and $x_2$, but $\xi_L^*$ does not.
However, for $\alpha=5$, $\xi_L^*$ almost has the reflection symmetry with respect to $x_2$.
The reflection symmetry with respect to $x_2$ can be easily verified by Theorem \ref{Th6}, but it is
not obvious with respect to $x_1$.  Note that Theorem \ref{Th6} only provides a sufficient condition for
the reflection symmetry.

\item   It takes less time to find $\xi_G^*$ than $\xi_L^*$, since the initial weight vector 
${\bf w}^{(0)}$ proposed in Remarks (ii) is very close to $\xi_G^*$ for the case of the GLSE in this example.
\end{enumerate}
Since there are only two design variables in the model, we can use a plot to show that $\xi_G^*$ and $\xi_L^*$ satisfy the conditition
(\ref{Cont2}). Let
$$  d(x_{1i}, x_{2i}) = \mbox{tr} \left( 2 {\bf G}^{-1}({\bf w}^{*}) G_i - {\bf H}^{-1}({\bf w}^{*}) H_i \right) - q, \quad i=1, \ldots, N,$$
where point $(x_{1i}, x_{2i})$ is the one used to evaluate the matrices $G_i$ and $H_i$.
Two representative plots are given in Figures \ref{Fig1} and \ref{Fig2},
and they are for $\xi_G^*$ with $\alpha=5$ and $\xi_L^*$ with $\alpha=0$, respectively.
It is clear from these plots that $  d(x_{1i}, x_{2i})$ are less than zero and the condition in
(\ref{Cont2}) is satisfied.

\begin{table}
	\begin{small}
	\caption{Minimax D-optimal designs and computation times in Example 2: all the support 
	points in $\xi_L^*$ are listed; only half of the  support 
	points in $\xi_G^*$ are listed and the other half are obtained by changing the sign of variable $x_1$. }
	\begin{center}
		\begin{tabular}{crrrrr} \hline
Case	 & \multicolumn{2}{c}{support points} & \multicolumn{3}{c}{weights} \\
	       &  $x_1$  & $x_2$ & $\alpha=0$  & $\alpha=3$ & $\alpha=5 $ \\ \hline \hline
	 $\xi_L^*$ & $-1 $ & $-1 $ & 0.1145 & 0.1145 & 0.1078  \\   
	           & $-1 $ & $1 $ & 0.0984 & 0.1003 & 0.1078  \\ 
	           & $-0.8$ & $0$ & 0.1430 & 0.1430 & 0.1389  \\ 
	           & $-0.3$ & $-1$ & 0 & 0 & 0.0651  \\   
	           & $-0.3$ & $0$ & 0 & 0 & 0.0157  \\ 
	           & $-0.3$ & $1$ & 0.1441 & 0.1422 & 0.0652  \\ 
	           & $0.3$ & $-1$ & 0.1441 & 0.1422 & 0.0730  \\   
	           & $0.3$ & $1$ & 0 & 0 & 0.0728  \\ 
	           & $0.8$ & $0$ & 0.1430 & 0.1430 & 0.1401  \\ 
	           & $1$ & $-1$ & 0.0984 & 0.1003 & 0.1068  \\ 
	           & $1$ & $1$ & 0.1145 & 0.1145 & 0.1068  \\  \hline
	    computation time (s)      &   &  & 235.8125   & 408.3109 & 1300.8017  \\
	    loss function $\tilde{\phi}_L$  &   &  & 58.2630  & 65.1178  &  68.1711 \\    \hline     \hline
	 $\xi_G^*$ & $-1$ & $-1$ & 0.0938 & 0.0806 & 0.0808  \\ 
	           & $-1$ & $0$ & 0.0336 & 0.0452 & 0.0448  \\ 
	           & $-1$ & $1$ & 0.0938 & 0.0806 & 0.0808  \\ 
	           & $-0.8$ & $-1$ & 0.0563 & 0.0511 & 0.0511  \\ 
	           & $-0.8$ & $0$ & 0.0291 & 0.0411 & 0.0411  \\ 
	           & $-0.8$ & $1$ & 0.0563 & 0.0511 & 0.0511  \\ 
	           & $-0.3$ & $-1$ & 0.0489 & 0.0459 & 0.0456  \\ 
	           & $-0.3$ & $0$ & 0.0393 & 0.0585 & 0.0591  \\ 
	           & $-0.3$ & $1$ & 0.0489 & 0.0459 & 0.0456  \\  \hline
	 computation time  (s)    &   &  & 46.0003 & 157.8750 & 163.4844  \\
	 loss function $\tilde{\phi}_G$  &   &  & 55.4642 & 63.7362 & 67.3218  \\     \hline    
	\end{tabular}  
	\end{center}
\label{table2}
\end{small}
\end{table}

\bigskip

{\bf Figure 1 here}

\bigskip

{\bf Figure 2 here}

There is another case that  $\xi_G^*$ and $\xi_L^*$  do not depend on ${\bf V}_0$ and $\alpha$.
In Theorem 4 of Wong et al. (2019), there is a result that the D-optimal design does not depend on
${\bf V}_0$ when $m=2$ and $ {\bf f}_1({\bf x})$ is a subvector of ${\bf f}_2({\bf x})$.
Following their proof, we can also show that 
$\xi_G^*$ and $\xi_L^*$  do not depend on ${\bf V}_0$ and $\alpha$
when $m=2$ and $ {\bf f}_1({\bf x})$ is a subvector of ${\bf f}_2({\bf x})$.
Since we can switch response varaibles in the model,
the result is also true if $ {\bf f}_2({\bf x})$ is a subvector of ${\bf f}_1({\bf x})$.
With Algorithm 1, we can explore a general result that 
$\xi_G^*$ and $\xi_L^*$  do not depend on ${\bf V}_0$ and $\alpha$ for  $m>2$ in Example 3.

\noindent{\bf Example 3.} Consider model \eqref{genlm-1} -- \eqref{Zmatrix} with $m=4$ and 3 design
variables, and 
\begin{eqnarray*}
	{\bf f}_1({\bf x})	&=&  (1, x_2, x_3)^\top, \\
	{\bf f}_2({\bf x})	&=&  (1, x_1, x_2, x_3, x_3^2)^\top, \\
	{\bf f}_3({\bf x})	&=&  (1, x_1, x_2, x_3, x_1x_3, x_3^2)^\top, \\
	{\bf f}_4({\bf x})	&=&  (1, x_1, x_2, x_3, x_1x_2, x_1x_3, x_2x_3,x_3^2)^\top, 
\end{eqnarray*}
where $ {\bf f}_1({\bf x})$ is a subvector of ${\bf f}_2({\bf x})$,
$ {\bf f}_2({\bf x})$ is a subvector of ${\bf f}_3({\bf x})$, and
$ {\bf f}_3({\bf x})$ is a subvector of ${\bf f}_4({\bf x})$, so that these vectors are 
``nested". 
Both $x_1$ and $x_2$ take 9 equally spaced points in $[0,1]$, 
$x_3$ takes 11 equally spaced points in $[-1, 1]$. 
and the design space $S_N$ has $N=9*9*11=891$ points.
$\xi_G^*$ and $\xi_L^*$ are computed for various ${\bf V}_0$ and $\alpha$, and our 
results show that 
$\xi_G^*$ and $\xi_L^*$ do not depend on ${\bf V}_0$ and $\alpha$. $\xi_G^*$ and $\xi_L^*$ are the same
for all the cases and they are given in Table \ref{table3}.  

\begin{table}
\begin{small}
		\caption{Minimax D-optimal design in Example 3 }
\begin{center}
\begin{tabular}{cccc} \hline
	 \multicolumn{3}{c}{support points} & {weights} \\
$x_1$  & $x_2$ & $x_3$ &  \\ \hline
0 & 0 & $-1$ & 0.0962 \\
0 & 0 & $0$ & 0.0576 \\
0 & 0 & $1$ & 0.0962 \\		
0 & 1 & $-1$ & 0.0962 \\
0 & 1 & $0$ & 0.0576 \\
0 & 1 & $1$ & 0.0962 \\		
1 & 0 & $-1$ & 0.0962 \\
1 & 0 & $0$ & 0.0576 \\
1 & 0 & $1$ & 0.0962 \\		
1 & 1 & $-1$ & 0.0962 \\
1 & 1 & $0$ & 0.0576 \\
1 & 1 & $1$ & 0.0962 \\		\hline		
\end{tabular}  
\end{center}
\label{table3}
\end{small}
\end{table}

\section{Conclusion}

We have investigated minimax D-optimal designs for multivariate regression models against
small departures of the assumed error matrix and obtained various analytical properties 
of the designs.  
In general it is hard to construct minimax designs analytically or numerically, since the objective function of minimax design problems 
is not convex.  However,
we are able to show  that the objective function of minimax D-optimal design problems is  a difference of two convex functions,
which makes the computation of minimax D-optimal designs tractable. We have developed an efficient and effective algorithm for 
finding  minimax D-optimal designs on discrete design spaces, and it is flexible to be applied for any 
 multivariate regression model. 
 
Minimax  D-optimal designs can be constructed based on the GLSE or the OLSE.  
How do we choose the estimator and $\alpha$ for practical applications? If we have an accurate estimate
${\bf V}_0$ of ${\bf V}_\epsilon$, then we use the GLSE to construct the minimax D-optimal design.  Otherwise, we can use the OLSE.
Since it is easy to compute minimax D-optimal designs  using Algorithm 1, it may be a good idea to do sensitivity 
analysis for the minimax D-optimal designs for various $\alpha$ values  and choose a minimax D-optimal design for a given application.

In this paper, we investigated the minimax D-optimality criterion. It would be interesting to study other minimax criteria, such as minimax A-optimality or minimax R-optimality. It is even more challenging to study other types of minimax optimal designs, as the objective functions are generally neither convex functions, nor difference of convex functions. 

Though we have focused on multivariate linear regression models, the methodology in this paper can be easily applied to nonlinear models for
finding locally minimax D-optimal designs.  In addition, we can apply the techniques in this paper to explore and construct  minimax D-optimal designs for regression models used in longitudinal studies (Chapter 4.2,  \citealt{diggle2002analysis}), where an outcome measure is taken from study participants at multiple time points.

\vspace{8cm}


\section*{Appendix: Proofs and derivations}

\noindent{\bf Proof of Theorem 1:}  From (\ref{M1M2}) and (\ref{Gmax}), we have
\begin{eqnarray*}
\phi_G(\xi, {\bf V}_0, \alpha) &=&	\max_{{\bf V}_{\epsilon} \in {\cal N}_\alpha({\bf V}_0)} \log \left(
\det({\bf M}_1(\xi, {\bf V}_0, {\bf V}_{\epsilon}) ) \right) \\
&=&	\max_{{\bf V}_{\epsilon} \in {\cal N}_\alpha({\bf V}_0)} \left( -2 
\log\left( \det({\bf A}(\xi, {\bf V}_0)) \right) + 
\log\left( \det({\bf B}(\xi, {\bf V}_0, {\bf V}_{\epsilon})) \right)  \right) \\
&=&	 -2 
\log\left( \det({\bf A}(\xi, {\bf V}_0)) \right) + 
\log\left( \det({\bf B}(\xi, {\bf V}_0, {\bf V}_0+\alpha {\bf I}_m)) \right),  ~~\mbox{by}~(\ref{N0eq})
\end{eqnarray*}
which gives the result in (\ref{phiG}).  The result in (\ref{phiL}) can be proved similarly. 
\hfill{$\Box$}

\bigskip

\noindent{\bf Proof of Lemma 1:} From (\ref{ABCD}), we obtain
\begin{eqnarray*}
{\bf B}(\xi, {\bf V}_0, {\bf V}_0+\alpha {\bf I}_m) &=& 
\sum_{i=1}^k w_i Z_i^\top {\bf V}_0^{-1} ({\bf V}_0+\alpha {\bf I}_m )  {\bf V}_0^{-1} Z_i	\\
&=& {\bf A}(\xi, {\bf V}_0) + \alpha \sum_{i=1}^k w_i Z_i^\top {\bf V}_0^{-2}Z_i	\\
& \succeq & {\bf A}(\xi, {\bf V}_0), ~\mbox{for all}~ \alpha \ge 0.
\end{eqnarray*}	
Thus, if ${\bf A}(\xi, {\bf V}_0)$ is nonsingular, then ${\bf B}(\xi, {\bf V}_0, {\bf V}_0+\alpha {\bf I}_m)$
is nonsingular.
The result about ${\bf D}(\xi, {\bf V}_0+\alpha {\bf I}_m)$ can be proved similarly.
\hfill{$\Box$}

\bigskip

\noindent{\bf Proof of Theorem 2:} 
From Theorem \ref{Th1}, we get
$$\phi_G(\xi_{\delta}, {\bf V}_0, \alpha)=-2 \log\left( \det({\bf A}(\xi_{\delta}, {\bf V}_0)) \right)
- (-\log\left( \det({\bf B}(\xi_{\delta}, {\bf V}_0, {\bf V}_0+\alpha {\bf I}_m)) \right) ).
$$
Since the weights of $\xi_{\delta}$ are linear in $\delta$,
from (\ref{ABCD}) it is easy to see that 
${\bf A}(\xi_{\delta}, {\bf V}_0)$  and 
${\bf B}(\xi_{\delta}, {\bf V}_0, {\bf V}_0+\alpha {\bf I}_m)$ are also linear in $\delta$.
By \citet[pg. 387]{boyd2004convex}, both 
$ - \log\left( \det({\bf A}(\xi_{\delta}, {\bf V}_0)) \right)$ and 
$- \log\left( \det({\bf B}(\xi_{\delta}, {\bf V}_0, {\bf V}_0+\alpha {\bf I}_m)) \right) $  are convex
functions of $\delta$, which implies the result for $\phi_G(\xi_{\delta}, {\bf V}_0, \alpha)$.
The result for $\phi_L(\xi_{\delta}, {\bf V}_0, \alpha)$ can be proved similarly.
\hfill{$\Box$}

\bigskip

\noindent{\bf Proof of Theorem 3:}  We prove the result for $\xi_G^*$. 
The proof for $\xi_L^*$  is similar and omitted.
On $S$, $\xi_G^*$ minimizes $\phi_G(\xi, {\bf V}_0, \alpha)$, and from Theorem \ref{Th1} we have
\begin{eqnarray*}
\phi_G(\xi, {\bf V}_0, \alpha) = -2 
\log\left( \det({\bf A}(\xi, {\bf V}_0)) \right) + 
\log\left( \det({\bf B}(\xi, {\bf V}_0, {\bf V}_0+\alpha {\bf I}_m)) \right).
\end{eqnarray*}	
For a scale transformation $T$, we define the design measure $\xi_T$ on $S_T$ with support points
$T{\bf x}_1, \ldots, T{\bf x}_k$ and their corresponding weights as $w_1, \ldots, w_k$.
Then on $S_T$, we minimize $\phi_G(\xi_T, {\bf V}_0, \alpha) = -2 
\log\left( \det({\bf A}(\xi_T, {\bf V}_0)) \right) + 
\log\left( \det({\bf B}(\xi_T, {\bf V}_0, {\bf V}_0+\alpha {\bf I}_m)) \right)$,
where, from  (\ref{ABCD}), (\ref{Zmatrix}) and the assumption in Theorem \ref{Th3},
\begin{eqnarray*}
&&{\bf A}(\xi_T, {\bf V}_0) =   {\bf Q}_T {\bf A}(\xi, {\bf V}_0) {\bf Q}_T, ~~\mbox{with~~}
{\bf Q}_T = Q_1 \oplus \ldots \oplus Q_m, \\
&& {\bf B}(\xi_T, {\bf V}_0, {\bf V}_0+\alpha {\bf I}_m) = {\bf Q}_T 
{\bf B}(\xi, {\bf V}_0, {\bf V}_0+\alpha {\bf I}_m) {\bf Q}_T. 	
\end{eqnarray*}
This gives
$$ \phi_G(\xi_T, {\bf V}_0, \alpha) = -\log \left( \det({\bf Q}_T) \right)^2 + 
\phi_G(\xi, {\bf V}_0, \alpha). $$
Since $Q_j$ do not depend on $w_1, \ldots, w_k$, minimizing 
$ \phi_G(\xi_T, {\bf V}_0, \alpha) $ 
over $w_1, \ldots, w_k$ is the same as minimizing   $ \phi_G(\xi, {\bf V}_0, \alpha) $.
Thus,  $\xi_G^*$ is scale invariant.
\hfill{$\Box$}

\bigskip

\noindent{\bf Proof of Theorem 4:}  
Define a $q \times q$ diagonal matrix $\tilde{\bf Q} = a_1 {\bf I}_{q_1} \oplus \ldots \oplus  a_m {\bf I}_{q_m}$,
where $a_1, \ldots, a_m$ are the diagonal elements of ${\bf Q}$,
and $q_j$ is the length of vector ${\bm \beta}_j$.  Notice that 
${\bf Q}^{-1} = {\bf Q}$ and $\det({\bf Q}) = \pm 1$.
If ${\bf V}_1={\bf Q} {\bf V}_0 {\bf Q}$, then from (\ref{ABCD}) we get
\begin{eqnarray*}
	{\bf A}(\xi, {\bf V}_1) &=& \sum_{i=1}^k w_i Z_i^\top {\bf V}_1^{-1} Z_i  \\
 		&=& \sum_{i=1}^k w_i Z_i^\top {\bf Q} {\bf V}_0^{-1} {\bf Q} Z_i \\
 		&=& \sum_{i=1}^k w_i \tilde{\bf Q}  Z_i^\top {\bf V}_0^{-1}  Z_i \tilde{\bf Q}, ~~~\mbox{using} ~(\ref{Zmatrix}) \\	
         &=& \tilde{\bf Q}  {\bf A}(\xi, {\bf V}_0)   \tilde{\bf Q},
\end{eqnarray*}
which gives that $\det\left( {\bf A}(\xi, {\bf V}_1) \right) =  \det\left( {\bf A}(\xi, {\bf V}_0) \right)$.  
Similarly we can show that 
$   \det\left( {\bf B}(\xi, {\bf V}_1, {\bf V}_1 + \alpha {\bf I}_m) \right) =  
\det\left( {\bf B}(\xi, {\bf V}_0, {\bf V}_0 + \alpha {\bf I}_m) \right)$
and  $\det\left( {\bf D}(\xi, {\bf V}_1 + \alpha {\bf I}_m) \right) =  
\det\left( {\bf D}(\xi, {\bf V}_0 + \alpha {\bf I}_m) \right)$.
Thus, by (\ref{phiG}) and (\ref{phiL}) we have 
$\phi_G(\xi, {\bf V}_1, \alpha) = \phi_G(\xi, {\bf V}_0, \alpha)$ and
$\phi_L(\xi, {\bf V}_1, \alpha) = \phi_L(\xi, {\bf V}_0, \alpha)$ for any $\xi$,
which implies the result in Theorem \ref{Th4}.
\hfill{$\Box$}

\bigskip

\noindent{\bf Proof of Theorem 5:} 
The objective function in the minimax D-optimal design problem (\ref{DC1})  can be written as 
$$g({\bf w}) - h({\bf w}) = -2 \log \left( \det( {\bf G}({\bf w})) \right) + 
\log \left( \det( {\bf H}({\bf w})) \right).$$
If ${\bf w}^*$ is a local minimizer, then it satisfies
that 
$$ \frac{\partial \left(g((1-\delta){\bf w}^* + \delta {\bf w}) - h((1-\delta){\bf w}^* + \delta {\bf w})\right) }
{\partial \delta} |_{\delta=0} \ge 0,$$
for any weight vector ${\bf w}$.  
Direct calculation of the above derivative gives 
$$  \mbox{tr} \left( 2 {\bf G}^{-1}({\bf w}^{*}) G_i - {\bf H}^{-1}({\bf w}^{*}) H_i \right) - q \le 0,
\mbox{~for~} i=1, \ldots, N.
$$
\hfill{$\Box$}

\bigskip
\noindent{\bf Proof of Theorem 6:} 
For  transformation $T_r$, we define the design measure $\xi_{T_r}$ on $S_{T_r}$ with support points
$T_r{\bf u}_1, \ldots, T_r{\bf u}_N$ and their corresponding weights as $w_1, \ldots, w_N$.
Following the proof of Theorem 3 and using the assumption in Theorem 6, 
we can show that the objective function 
$g({\bf w}) -v({\bf w}, {\bf w}^0)$ in problem (\ref{DC1A})  is the same for the 
design measures  $\xi$ on $S_{N}$  and $\xi_{T_r}$ on $S_{T_r}$.
Notice that the convex combination of $\xi$ and $\xi_{T_r}$, $ 0.5 \xi + 0.5 \xi_{T_r} $,
has the reflection symmetry with respect to
variable $x_r$.
Since $g({\bf w}) -v({\bf w}, {\bf w}^0)$  is a convex function of ${\bf w}$, it is clear that 
there exists a solution to problem (\ref{DC1A}) that has the reflection symmetry with respect to
variable $x_r$.  
In Algorithm 1, there exist a sequence of ${\bf w}^{(l)}$, $l=1, 2, \ldots$, that have 
the reflection symmetry with respect to
variable $x_r$.  This implies that the limit ${\bf w}^*$ of ${\bf w}^{(l)}$ as $l \to \infty$ also has
the reflection symmetry.
\hfill{$\Box$}

\bigskip

\noindent{\bf Closed form formula for the gradient $\nabla h({\bf w})$:} The $i$th element of $\nabla h({\bf w})$ is given by
$$ - \mbox{tr} \left( {\bf H}^{-1}({\bf w}) H_i \right),$$
where matrices ${\bf H}({\bf w})$ and $ H_i$ are defined in Theorem \ref{Th5}.
\hfill{$\Box$}

\bigskip

\section*{Acknowledgements}

This research work is partially supported by Discovery  Grants from the Natural
Science and Engineering Research Council of Canada.

\bibliography{ref}

\end{document}